\documentclass[reprint, aps,prl,showkeys]{revtex4-1}

\usepackage{graphicx,epsfig}
\usepackage{amssymb}
\usepackage{amsmath}
\usepackage{bm}
\usepackage{textcomp}
\usepackage[utf8]{inputenc}
\usepackage{color}

\begin{document}
\setcounter{page}{1}

\title[]{Signatures of an annular Fermi sea}
\author{Insun \surname{Jo}}
\author{Yang \surname{Liu}}
\author{L. N. \surname{Pfeiffer}}
\author{K. W. \surname{West}}
\author{K. W. \surname{Baldwin}}
\author{M. \surname{Shayegan}}
\affiliation{Department of Electrical Engineering, Princeton University, Princeton, NJ 08544, USA  }
\author{R. \surname{Winkler}}
\affiliation{Department of Physics, Northern Illinois University, Dekalb, IL 60115, USA   }

\begin{abstract}
  We report Shubnikov-de Haas oscillations measurements revealing experimental
  signatures of an annular Fermi sea that develops near the energy
  band edge of the excited subband of two-dimensional holes confined
  in a wide GaAs quantum well. As we increase the hole density, when
  the Fermi level reaches the excited subband edge, the low-field
  magnetoresistance traces show a sudden emergence of new oscillations at an unexpectedly large
  frequency whose value does \textit{not} correspond to the
  (negligible) density of holes in the excited subband. There is
  also a sharp and significant increase in zero-field resistance
  near this onset of subband occupation. Guided by numerical energy
  dispersion calculations, we associate these observations with the
  unusual shape of the excited subband dispersion which results in a
  ``ring of extrema'' at finite wavevectors and an annular Fermi
  sea. Such a dispersion and Fermi sea have long been expected from
  energy band calculations in systems with strong spin-orbit
  interaction but their experimental signatures have been elusive.
\end{abstract}

\maketitle

The concept of a Fermi surface, the constant-energy surface containing all the occupied electron states in momentum, or wavevector ($k$), space plays a key role in determining electronic properties of conductors \cite{AshMerin}. The connectivity and topology of the Fermi surface have long been of great interest \cite{Lifshitz.JETP.1960}. In two-dimensional (2D) carrier systems, the Fermi surface becomes a ``contour'' which, in the simplest case, encircles the occupied states (see Fig. 1(a)). In this case, the area inside the contour, which we refer to as the Fermi sea (FS), is a simple disk. In 2D systems with multiple conduction band valleys, e.g. 2D electrons confined to Si or AlAs quantum wells (QWs) \cite{Takashina.PRL.2006,Shkolnikov.PRL.2002,Gunawan.PRL.2004} or 2D electrons in a wide GaAs QW subject to very large parallel fields \cite{Mueed.PRL.2015}, the FS consists of a number of separate sections, each containing a fraction of the electrons in the system (Fig. 1(b)). Figure 1(c) shows yet another possible FS topology, namely an annulus. Such a FS is expected in systems with a strong Rashba spin-orbit interaction (SOI) \cite{Vasko.SPSS.1979,Bychkov.JETPL.1984,RolandBook,Cappelluti.PRL.2007,Cheuk.PRL.2012,Bihlmayer.NJP.2015,Brosco.PRL.2016,Nichele.Arxiv.2016}, biased bilayer graphene \cite{McCann.PRL.2006,Castro.PRL.2008,Varlet.PRL.2014,Stauber.PRB.2007}, or monolayer gallium chalcogenides \cite{Wu.Arxiv.2014,Cao.PRL.2015}. Since the electron states near the band extremum become highly degenerate, resulting in a van Hove singularity in the density of states, an annular FS has been predicted to host exotic
interaction-induced phenomena and phases such as ferromagnetism \cite{Stauber.PRB.2007,Wu.Arxiv.2014,Cao.PRL.2015}, anisotropic Wigner crystal and nematic phases \cite{Berg.PRB.2012, Ruhman.PRB.2014,fnote1,Kernreiter.PRB.2013}, and a persistent current state \cite{Jung.PRB.2015}.

\begin{figure}
  \begin{center}
    \psfig{file=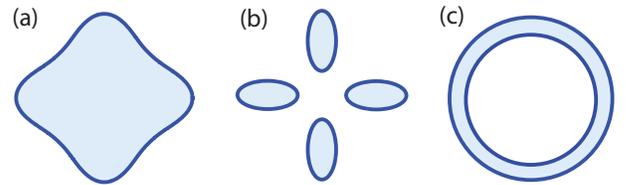, width=0.45\textwidth }
  \end{center}
  \caption{\label{fig1} Examples of Fermi seas in 2D systems: (a) simple disk, (b) multi-fold ellipses, (c) annulus. }
\end{figure}

Although the possibility of an annular FS has long been recognized theoretically, its direct detection has been elusive. For cases (a) and (b) in Fig. 1, the FS can readily be probed as the frequencies of the Shubnikov-de Haas (SdH) oscillations, multiplied by $e/h$, directly give the FS area or, equivalently, the areal density of the 2D system \cite{Onsager.PM.1952,Shkolnikov.PRL.2002,Gunawan.PRL.2004,Liu.PRB.2011,Mueed.PRL.2015,fnote2,Winkler.PRL.2000} ($e$ is electron charge and $h$ is the Planck constant). For the annular FS of case (c), however, it is not known how the frequencies of the oscillations are related to the area of the FS. Here we report energy band calculations and experimental data demonstrating the observation of an annular FS and its unusual SdH oscillations in 2D hole systems (2DHSs) confined in wide GaAs QWs. 

\begin{figure}
  \begin{center}
    \psfig{file=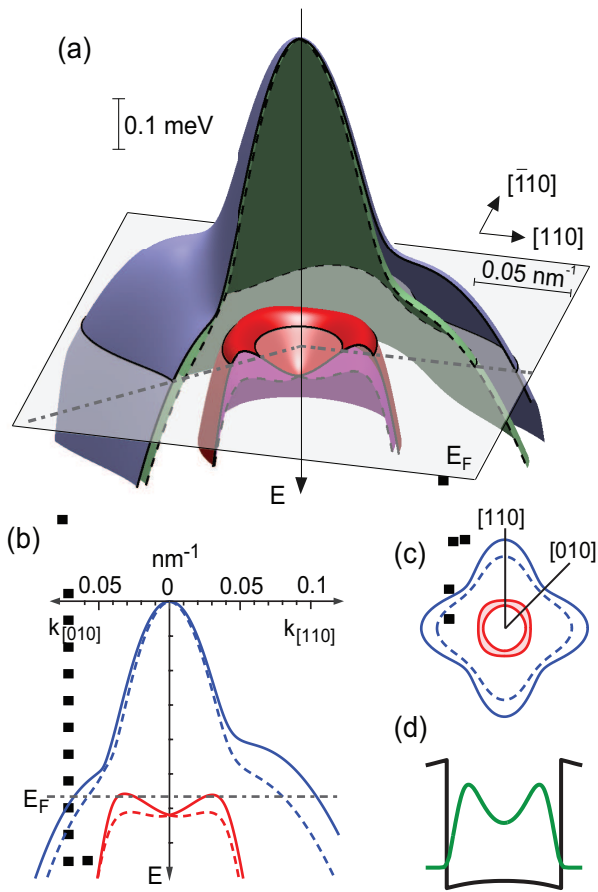, width=0.5\textwidth }
  \end{center}
  \caption{\label{fig2} (a) and (b) Calculated subband energy dispersions for GaAs 2D holes in a 38-nm-wide GaAs QW at density $p=1.2\times 10^{11}$~cm$^{-2}$. The upper (lower) dispersions are for the ground (excited) subbands, each consisting of two spin-split branches at finite $k$, shown by solid and dashed lines. The horizontally-cut plane in (a) and the dash-dotted line in (b) mark the Fermi energy. In (b) the energy dispersions are shown along two directions. (c) Fermi contours for the ground (blue) and excited (red) subband. The FSs for the ground subband are not filled with color for clarity; the FS for the excited subband is the red annulus. (d) Charge distribution and potential.}
\end{figure}

Figure 2 highlights the key points of our study. Figures 2(a) and 2(b) show
the calculated energy band dispersions for a 2DHS at a density of
$p=1.2\times 10^{11}$~cm$^{-2}$ confined in a 38-nm-wide GaAs QW
\cite{fnote3}. The self-consistent calculations are based on the $8\times 8$
Kane Hamiltonian \cite{RolandBook}. The charge distribution is
bilayer-like (Fig. 2(d)) because the Coulomb repulsion pushes the carriers towards the confinement walls
\cite{Suen.PRB.1991, Suen.PRL.1992, Liu.PRL.2014}. As seen in
Figs. 2(a) and (b), the energy band dispersion is very unusual,
showing an \textit{inverted} structure for the
excited subband with a ``ring of maxima'' at finite values of $k$. So far, such dispersions have been studied mostly within systems with Rashba SOI \cite{Vasko.SPSS.1979,Bychkov.JETPL.1984,RolandBook,Cappelluti.PRL.2007,Cheuk.PRL.2012,Bihlmayer.NJP.2015,Brosco.PRL.2016}. However, in our symmetric 2DHS (without the Rashba SOI), the inverted band structure stems from the combined effect of a strong level repulsion between the second heavy-hole and the first light-hole subbands at $k>0$ \cite{Bastard.1988} as well as the Dresselhaus SOI \cite{Dresselhaus.PR.1955}.
When holes start to occupy this excited subband, its FS adopts an annular shape (Fig. 2(c)). Unlike the
FS of the ground subband, the annular FS has a
void for small $k$. In our experiments we probe the energy band
dispersions via measuring SdH oscillations as a function of
increasing density. We observe a sudden appearance of an extra peak
at a relatively large frequency in the Fourier transform (FT)
spectra as the holes start to occupy the excited subband. We
associate this peak with the annular FS, and discuss the
details of its evolution with increasing hole density.

\begin{figure*}
  \begin{center}
    \psfig{file=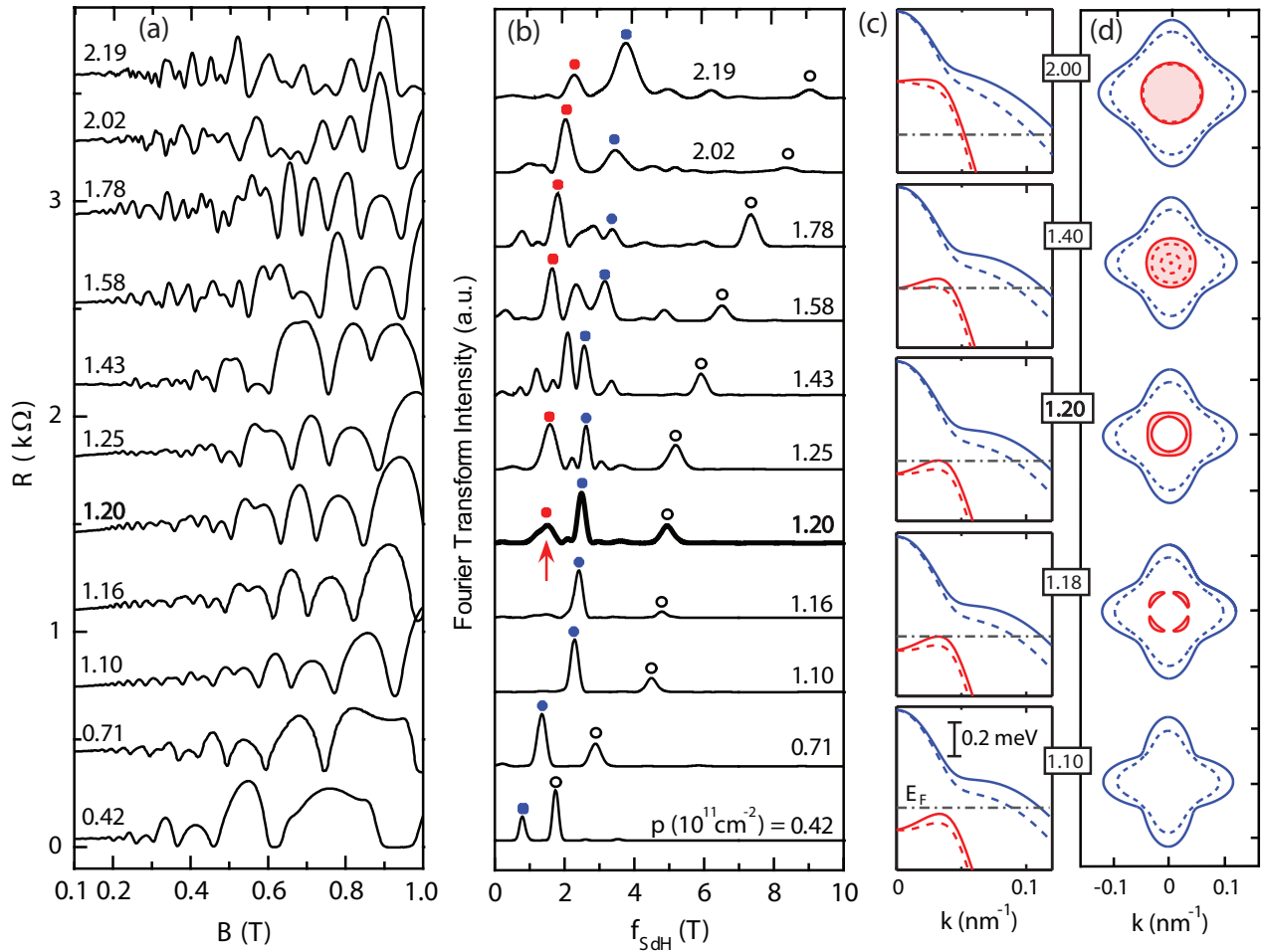, width=0.95 \textwidth }
  \end{center}
  \caption{\label{fig3} (a) Low-field magnetoresistance traces at
    different densities. (b) Fourier transform spectra of the SdH oscillations at each density. Open
    circles indicate total density peaks. Blue and red circles mark the peaks associated with the ground and excited subbands (see text). (c) Calculated energy dispersions and, (d) Fermi sea, at densities $p=1.10, 1.18, 1.20, 1.40, 2.00~\times 10^{11}$ cm$^{-2}$, from bottom to top. Gray
    dash-dotted lines in (c) represent the Fermi energy. Blue and red lines
    correspond to the ground and excited subbands, respectively; solid and dashed lines
    represent the spin-split states.}
\end{figure*}

In our experiments, we used 40-nm-wide, symmetric, GaAs QWs grown by
molecular beam epitaxy along the [001] crystal direction \cite{fnote3}. The QWs
are symmetrically modulation doped with two C~$\delta$-layers. We used two samples with different Al$_{0.3}$Ga$_{0.7}$As spacer-layer thicknesses (160 and 90 nm), different as-grown hole densities ($p=1.1$ and 2.0, in units of $10^{11}$ cm$^{-2}$ which we use throughout this manuscript), and mobilities (32 and 76 m$^{2}$/Vs). The data for $p\leq1.43$ are taken from the lower density sample, and the higher density data from the other sample. In each sample, front- and back-gate electrodes allow us to change independently the 2D hole density and the
asymmetry of the charge distribution in the QW. In this study, we focus on symmetric charge
distributions; we judge the symmetry via a careful examination of
the SdH oscillations in the low-density regime where only the
ground subband is occupied \cite{fnote4,
  Papadakis.Science.1999}, as well as the strengths of fractional
quantum Hall states, e.g. at $\nu=1/2$ \cite{Liu.PRL.2014}. The
low-field magnetoresistance oscillations are measured in a dilution
refrigerator at a base temperature of $\sim$50 mK.

Figure 3 shows the evolutions of: (a) low-field magnetoresistance data, (b) their corresponding FT spectra, (c) the calculated energy dispersions, and (d) the associated Fermi contours. For clarity, the traces in Fig. 3(a) and their FTs in Fig. 3(b) are shifted vertically. The intensities of the FTs are normalized so that the heights of the strongest FT peaks in different spectra are comparable. For all traces the total density is determined from the position of the high-frequency peak that is marked by an empty circle following Onsager \cite{Onsager.PM.1952}, i.e., by multiplying the frequency by $e/h$. This density agrees well, to within 4\%, with the magnetic field positions of the integer and fractional quantum Hall states observed at high fields. At low densities ($p<1.2$) the FT spectra are simple. Besides a peak corresponding to the total density, we also observe a second peak, marked by a blue closed circle, at half value of the total density peak. We associate this peak with SdH oscillations at very low magnetic fields where the Zeeman energy is small and the spin splitting of the Landau levels of the ground subband is not yet resolved.

As the density is raised to $p=1.20$, a FT peak suddenly appears at $\simeq$1.5 T in Fig. 3(b). This peak, which is marked by a red arrow and circle in the $p=1.20$ trace, signals the onset of the excited-subband occupation. As we discuss later (Fig. 4), there is also a rather sharp rise in the sample resistance at $p=1.20$, consistent with our conjecture. This new peak has two unusual characteristics. First, its emergence is very abrupt. It is essentially absent at a slightly lower density of $p=1.16$, and its strength grows very quickly to become the dominant peak in the whole FT spectrum at a slightly higher density of $p=1.25$. Second, its frequency, multiplied by the usual factors ($e/h$ or $2e/h$), clearly does \textit{not} give the correct density of holes in the excited subband, which we expect to be extremely small, essentially zero. Consistent with this expectation, the peak near 2.5 T (marked by a blue circle), which we associate with the ground subband, indeed accounts for essentially \textit{all} the QW's holes: 2.5 T multiplied by $2e/h$ gives $p=1.20$, leaving very few holes for the excited subband. We conclude that the frequency of the $f\simeq1.5$ T peak is not related to the excited-subband density. This is in sharp contrast to the GaAs 2D \textit{electron} systems where, after the onset of the excited-subband occupation, a FT peak appears at a small frequency and the frequency slowly increases as more electrons occupy the excited subband \cite{Stormer.SSC.1982,Stormer.JVST.1982,Shayegan.PRL.1990, Lu.PRB.1998, Lu.Thesis.1998}. Moreover, the frequency of the low-frequency peak correctly gives the electron density of the excited subband.

We associate the $f\simeq1.5$ T peak with the onset of the excited-subband occupation and the emergence of an annular FS in our 2DHS. But, how should an annular FS be manifested in SdH oscillations? Given that the frequency of this peak does not correspond to the area of the annulus, is there an alternative relation? Following Onsager \cite{Onsager.PM.1952}, one may speculate that it could lead to oscillations whose frequencies are given by the areas enclosed by the outer and inner circles (or more generally, the ``contours'') of the annulus, namely by the areas $\pi k_o^2$ and $\pi k_i^2$, where $k_o$ and $k_i$ are the radii of the outer and inner circles. Near the onset of the excited-subband occupation, the outer and inner contours of the annulus are very close to each other \cite{fnote5}. We would then expect the FT to show two closely-spaced peaks. Based on the energy band calculations, for $p=1.20$, we expect FT peaks at $f=0.43$ T and $f=0.27$ T for the outer and inner rings, respectively. These values are smaller than the frequency ($f\simeq1.5$ T) of the broad peak we observe in the FT. The discrepancy might imply that this speculation is not entirely correct, or that the band calculations are not quantitatively accurate.

The evolution of the FT spectra for $p>1.25$ is also suggestive. For
$p=1.43$ the spectrum becomes quite complex, showing multiple peaks
near 2 T. This is qualitatively consistent with the results
of the energy band calculations: Near the onset of the excited-subband
occupation, $E_{\textrm{F}}$ can have four crossings with the
excited-subband dispersion (two for each spin-subband dispersion),
resulting in two, complex annular FSs (see Figs. 3(c) and (d)
for $p=1.40$). As we further increase the density, the FT spectra
become simpler, showing two dominant peaks at the highest densities
(see FTs for $p=2.02$ and 2.19 in Fig. 3(b)). Such an evolution
qualitatively agrees with our expectation based on the calculated
bands which indicate two ``normal'' FSs (i.e., without voids
at $k=0$), one for each subband. Also consistent with calculations,
these peaks move to higher frequencies when the densities of
subbands increase with increasing total density. If we assign these
peaks to the areas of the FSs for the ground and
excited subbands, multiply their frequencies by $2e/h$, and sum the
two densities, we find a total density which is $\sim30\%$ larger than
the total density expected from the open circles. If we assume that the
excited-subband Landau levels are spin-resolved, and multiply the
lower frequency (red) peak by $e/h$ (instead of $2e/h$), then we
obtain a total density which agrees to better than $\sim8\%$ with
the total density deduced from the open circles.

\begin{figure}
  \begin{center}
    \psfig{file=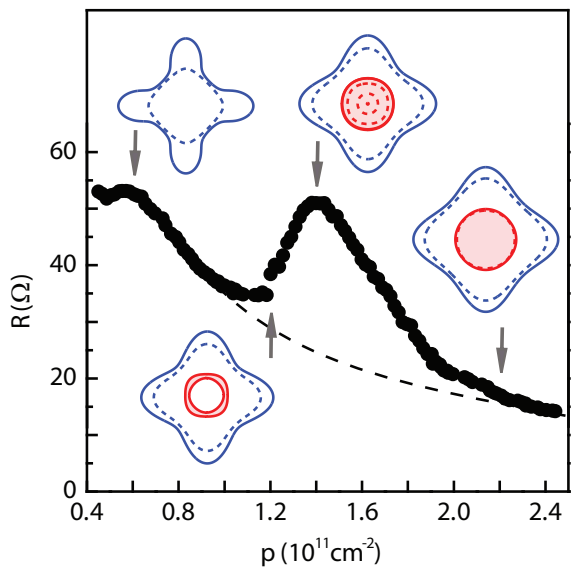, width=0.45\textwidth }
  \end{center}
  \caption{\label{fig4} Zero-field resistance as a
    function of density. Fermi seas are shown for four densities indicated by arrows.}
\end{figure}

We also measured the zero-field resistance of the 2DHS as
a function of the total density in the QW, and show the results in
Fig. 4. The data corroborate our conclusions regarding the
occupation of the excited subband. At low densities ($p<1.2$) where
only the ground subband is occupied, the resistance decreases
with increasing density, consistent with an increase in conductivity
because of the larger density and higher mobility. At $p=1.2$, the
resistance shows an abrupt increase. A qualitatively similar rise is
also seen at the onset of the occupation of the excited subband in
GaAs 2D electrons \cite{Stormer.JVST.1982,Stormer.SSC.1982,
  Lu.Thesis.1998}, and can be attributed to the enhanced
inter-subband scattering. In Fig. 4, the resistance increases with
increasing density past $p=1.2$ and attains a maximum value for
$p\simeq1.4$, the density near which we see multiple FT peaks
(Fig. 3(b)). This is consistent with extra scattering among the
multiple branches of the hole subbands (Fig. 3(c)). When $p$ is further increased, the resistance decreases monotonically. Again
this agrees with the calculated dispersions; for $p>1.4$, $E_{\textrm{F}}$ goes past the inverted band structure and the holes continue
to occupy both subbands, thus decreasing the resistance.

In conclusion, our study of low-field SdH oscillations for 2D holes confined in a wide QW reveals signatures of an annular FS that originates from the inverted dispersion of the excited subband. When the excited subband begins to be populated, in the FT spectrum we observe a sudden emergence of an anomalous peak whose frequency is not associated with the density of holes in the excited subband through the usual Onsager relation. We add that near the onset of this population, the holes in the excited subband occupy only one spin branch of the dispersion. This is qualitatively different from the usual case (e.g., the ground subband) where, in the absence of the linear-$k$ SOI, the holes occupy both spin-subbands even at the onset of the occupation. Our results should stimulate future experimental and theoretical studies of the unusual dispersion and annular FS.

\begin{acknowledgments}
  We acknowledge support by the DOE BES (DE-FG02-00-ER45841) grant
  for measurements, and the NSF (Grants DMR-1305691, DMR-1310199 and MRSEC
  DMR-1420541), the Gordon and Betty Moore Foundation (Grant
  GBMF4420), and Keck Foundation for sample fabrication and
  characterization. We also acknowledge support by the NSF for
  low-temperature equipment (Grant DMR-MRI-1126061).
  We appreciate stimulating discussions with D.~Culcer, R.~J.\ Joynt, J. Jung, 
  L.~Smrcka, K.~Vyborny, and U.~Z\"ulicke. 
\end{acknowledgments}


\begin{thebibliography}{99}
\bibitem{fnote3} Our samples' nominal QW width is 40 nm. In our calculations, we use a width of 38 nm so that the onset of the excited subband occupation, $p=1.2 \times 10^{11}$~cm$^{-2}$, matches the experimental value.
\bibitem{fnote5} Right at the onset of the excited-subband occupation and in an extremely narrow density range near $p=1.18$, the calculated FS of the excited-subband consists of four ``arcs'' (see Fig. 3(d)). We do not seem to observe a signature of such a FS in SdH measurements, likely because of the extremely small density of holes in the arcs, and also because of the very narrow density range where the arcs prevail.
\end{thebibliography}
\end{document}